\documentclass[sigconf]{acmart}
\usepackage{graphicx}
\usepackage{algorithm}
\usepackage[noend]{algpseudocode}
\usepackage{algorithmicx}
\usepackage{textcomp}
\usepackage{xcolor}
\usepackage{calc}        % for \dimexpr
\usepackage{amsmath,amsfonts}
\usepackage{marvosym}
\usepackage{listings}
\usepackage{enumitem}
\usepackage{tikz}
\usetikzlibrary{shapes,backgrounds}
\usepackage{framed}
\usepackage{float}
\usepackage{multirow}
\usepackage{array}
\usepackage{xspace}
\usepackage[breakable]{tcolorbox}
\hypersetup{colorlinks=true,linkcolor=blue,citecolor=blue}
\usepackage{cleveref}
\usepackage{comment}
\usepackage{url}
\usepackage{booktabs}
\newcommand{\approach}{\textsc{LLM2Ltac}}
\newcommand{\coqart}{\textsc{Coq-Art}}
\newcommand{\extlib}{\textsc{Ext-Lib}}
\newcommand{\vfa}{\textsc{Vfa}}
\newcommand{\compcert}{\textsc{CompCert}}
\newcommand{\coqcorn}{\textsc{C-CoRN}}

\newcommand{\theoremsum}{6199}
\newcommand{\finalres}{23.87\%}
\newcommand{\finalresagent}{9.90\%}
\newcommand{\finalrescost}{10.51\%}

\newtcolorbox{rqbox}{breakable,left=4pt,right=4pt,top=4pt,bottom=4pt}
\newcommand{\stratrocq}{\textsc{Strat2Rocq}}

\usepackage{xcolor}
\usepackage{listings}
\newcommand\code[1]{{\tt\small #1}}
\definecolor{dkgreen}{rgb}{0,0.3,0}
\definecolor{cmtgreen}{rgb}{0,0.6,0}
\definecolor{ltblue}{rgb}{0,0.4,0.4}
\definecolor{dkviolet}{rgb}{0.3,0,0.5}
\definecolor{dkblue}{rgb}{0,0.2,0.2}
\definecolor{dkred}{rgb}{0.6,0,0}
\lstdefinelanguage{Coq}{ 
    mathescape=true,
    texcl=false, 
    escapeinside={(@}{@)},
    morekeywords=[1]{Section, Module, End, Require, Import, Export,
        Variable, Variables, Parameter, Parameters, Axiom, Hypothesis,
        Hypotheses, Notation, Local, Tactic, Reserved, Scope, Open, Close,
        Bind, Delimit, Definition, Let, Ltac, Fixpoint, CoFixpoint, Add,
        Morphism, Relation, Implicit, Arguments, Unset, Contextual,
        Strict, Prenex, Implicits, Inductive, CoInductive, Record,
        Structure, Canonical, Coercion, Context, Class, Global, Instance,
        Program, Infix, Theorem, Lemma, Corollary, Proposition, Fact,
        Remark, Example, Proof, Goal, Save, Qed, Defined, Hint, Resolve,
        Rewrite, View, Search, Show, Print, Printing, All, Eval, Check,
        Projections, inside, outside, Def},
    morekeywords=[2]{forall, exists, exists2, fun, fix, cofix, struct,
        match, with, end, as, in, return, let, if, is, then, else, for, of,
        nosimpl, when},
    morekeywords=[3]{Type, Prop, Set, true, false, option},
    morekeywords=[4]{pose, set, move, case, elim, apply, clear, hnf,
        intro, intros, generalize, rename, pattern, after, destruct,
        induction, using, refine, inversion, injection, rewrite, congr,
        unlock, compute, ring, field, fourier, replace, fold, unfold,
        change, cutrewrite, simpl, have, suff, wlog, suffices, without,
        loss, nat_norm, assert, cut, trivial, revert, bool_congr, nat_congr,
        symmetry, transitivity, auto, split, left, right, autorewrite},
    morekeywords=[5]{by, done, exact, reflexivity, tauto, romega, omega,
        assumption, solve, contradiction, discriminate},
    morekeywords=[6]{do, last, first, try, idtac, repeat},
    morecomment=[s]{(*}{*)},
    showstringspaces=false,
    morestring=[b]",
    morestring=[d]’,
    tabsize=3,
    extendedchars=false,
    sensitive=true,
    breaklines=false,
    basicstyle=\fontsize{9.5pt}{11.4pt}\selectfont\ttfamily,
    captionpos=b,
    columns=[l]flexible,
    identifierstyle={\ttfamily\color{black}},
    keywordstyle=[1]{\bfseries\ttfamily\color{dkviolet}},
    keywordstyle=[2]{\bfseries\ttfamily\color{dkgreen}},
    keywordstyle=[3]{\bfseries\ttfamily\color{ltblue}},
    keywordstyle=[4]{\bfseries\ttfamily\color{dkblue}},
    keywordstyle=[5]{\bfseries\ttfamily\color{dkred}},
    stringstyle=\ttfamily,
    commentstyle={\bfseries\ttfamily\color{cmtgreen}},
    literate=
    {\\forall}{{\color{dkgreen}{$\forall\;$}}}1
    {\\exists}{{$\exists\;$}}1
    {<-}{{$\leftarrow\;$}}1
    {=>}{{$\Rightarrow\;$}}1
    {==}{{\code{==}\;}}1
    {==>}{{\code{==>}\;}}1
    {->}{{$\rightarrow\;$}}1
    {<->}{{$\leftrightarrow\;$}}1
    {<==}{{$\leq\;$}}1
    {\#}{{$^\star$}}1 
    {\\o}{{$\circ\;$}}1 
    {\@}{{$\cdot$}}1 
    {\/\\}{{$\wedge\;$}}1
    {\\\/}{{$\vee\;$}}1
    {++}{{\code{++}}}1
    {~}{{$\sim$}}1
    {\@\@}{{$@$}}1
    {\\mapsto}{{$\mapsto\;$}}1
    {\\hline}{{\rule{\linewidth}{0.5pt}}}1
}[keywords,comments,strings]

\lstdefinelanguage{PlainText}{ 
    mathescape=true,
    texcl=false, 
    escapeinside={(@}{@)},
    morecomment=[s]{(*}{*)},
    showstringspaces=false,
    morestring=[b]",
    morestring=[d]’,
    tabsize=3,
    extendedchars=false,
    sensitive=true,
    breaklines=false,
    basicstyle=\fontsize{5.5pt}{7.5pt}\selectfont\ttfamily,
    captionpos=b,
    columns=[l]flexible,
    identifierstyle={\ttfamily\color{black}},
    keywordstyle=[1]{\bfseries\ttfamily\color{dkviolet}},
    keywordstyle=[2]{\bfseries\ttfamily\color{dkgreen}},
    keywordstyle=[3]{\bfseries\ttfamily\color{ltblue}},
    keywordstyle=[4]{\bfseries\ttfamily\color{dkblue}},
    keywordstyle=[5]{\bfseries\ttfamily\color{dkred}},
    stringstyle=\ttfamily,
    commentstyle={\bfseries\ttfamily\color{dkgreen}},
    literate=
    {\\forall}{{\color{dkgreen}{$\forall\;$}}}1
    {\\exists}{{$\exists\;$}}1
    {<-}{{$\leftarrow\;$}}1
    {=>}{{$\Rightarrow\;$}}1
    {==}{{\code{==}\;}}1
    {==>}{{\code{==>}\;}}1
    {->}{{$\rightarrow\;$}}1
    {<->}{{$\leftrightarrow\;$}}1
    {<==}{{$\leq\;$}}1
    {\#}{{$^\star$}}1 
    {\\o}{{$\circ\;$}}1 
    {\@}{{$\cdot$}}1 
    {\/\\}{{$\wedge\;$}}1
    {\\\/}{{$\vee\;$}}1
    {++}{{\code{++}}}1
    {~}{{$\sim$}}1
    {\@\@}{{$@$}}1
    {\\mapsto}{{$\mapsto\;$}}1
    {\\hline}{{\rule{\linewidth}{0.5pt}}}1
}[keywords,comments,strings]

\algrenewcommand\algorithmicrequire{\textbf{Input:}}
\algrenewcommand\algorithmicensure{\textbf{Output:}}
\AtBeginDocument{%
  }

\setcopyright{acmlicensed}
\copyrightyear{2018}
\acmYear{2018}
\acmDOI{XXXXXXX.XXXXXXX}
\acmConference[Conference acronym 'XX]{Make sure to enter the correct
  conference title from your rights confirmation email}{June 03--05,
  2018}{Woodstock, NY}
\acmISBN{978-1-4503-XXXX-X/2018/06}

\newcommand{\yxmodifyok}[2]{{#2}}

\begin{document}

\title{A Learning Method for Symbolic Systems Using Large Language Models}


\author{Jian Fang}
\affiliation{%
  \institution{Key Laboratory of High Confidence Software Technologies (Peking University), Ministry of Education; School of Computer Science, Peking University}
  \city{Beijing}
  \country{China}
}
\email{fangjian@stu.pku.edu.cn}

\author{Yixun Yao}
\affiliation{%
  \institution{Key Laboratory of High Confidence Software Technologies (Peking University), Ministry of Education; School of Computer Science, Peking University}
  \city{Beijing}
  \country{China}
}
\email{yaoyixun@stu.pku.edu.cn}

\author{Yingfei Xiong\textsuperscript{\Letter}}
\affiliation{%
  \institution{Key Laboratory of High Confidence Software Technologies (Peking University), Ministry of Education; School of Computer Science, Peking University}
  \city{Beijing}
  \country{China}
}
\email{xiongyf@pku.edu.cn}


\begin{abstract}
% Automated theorem proving is important for formal verification of safety-critical systems. Symbolic provers are complementary to large language models (LLMs) in terms of proving capability and do not require expensive  computation resources, and thus improving the performance of symbolic provers is important. 

% This paper proposes \approach{}, the first approach that utilizes LLMs to learn new tactics from existing proofs to enhance symbolic provers. Given a corpus of formal proofs, \approach{} first asks an LLM to identify reusable proof strategies from the proofs and formalize them as tactics. These tactics are verified to be valid and generalizable, and finally integrated with symbolic provers to enhance their automated proving capabilities.

Automated theorem proving is essential for the formal verification of safety-critical systems. As the corpus of formal proofs grows, a natural paradigm is to learn from existing proofs. However, current learning-based approaches predominantly train Large Language Models (LLMs) as end-to-end provers, which yields resource-intensive, opaque systems. 
Conversely, while traditional symbolic provers are computationally efficient, how to automatically improve these solvers from data remains an open challenge.
% While learning reasoning strategies directly from data is highly desirable, how to automatically learn such a system remains an open challenge.
% While learning a lightweight symbolic prover directly from data is highly desirable, how to automatically learn such a system remains an open challenge.

This paper bridges this gap by proposing \approach{}, the first approach that leverages the reasoning power of LLMs not as end-to-end provers, but as intelligent synthesizers to mine purely symbolic tactics from data. Given a corpus of formal proofs, \approach{} asks an LLM to identify latent proof strategies and formalize them into reusable tactics. These tactics are verified for validity and generalizability, and finally integrated into symbolic provers to enhance their automated proving capabilities without the runtime cost of LLMs.

We implement \approach{} on Rocq 8.20.0 and mine tactics from 11,725 theorems in the standard library. %We test the learned tactics at 93,424 positions across 6,462 theorems from an independent project to ensure they generalize beyond the extraction dataset.
We evaluate our approach on 6,199 theorems from four large real-world verification projects, namely, \compcert{}, \coqart{}, \extlib{}, and \vfa{}. 
Results show that the mined tactics improve CoqHammer to prove \finalres{} more theorems, and when integrating the improved CoqHammer with Claude Code, the overall proved theorems increases by \finalresagent{},
indicating the effectiveness of \approach{}.
\end{abstract}




\maketitle

\section{Introduction}
\label{sec:intro}
Automating formal verification is important for ensuring the correctness of safety-critical software~\cite{Leroy-BKSPF-2016,DBLP:conf/sosp/KleinEHACDEEKNSTW09,DBLP:conf/osdi/GuSCWKSC16} and hardware systems~\cite{DBLP:conf/cav/SuYCLBH25,rIC3}. Interactive theorem provers (ITPs) such as Rocq (Coq)~\cite{CoqRefMan8.20}, Isabelle~\cite{DBLP:conf/tphol/WenzelPN08}, and Lean~\cite{DBLP:conf/cade/Moura021} provide rigorous mathematical frameworks for constructing machine-checked proofs. However, manual proof construction requires substantial expertise and effort, significantly limiting the scalability of formal verification to large-scale industrial applications.

The state-of-the-art approach to automating formal verification combines large language models (LLMs) with symbolic automated theorem provers (ATPs)~\cite{DBLP:conf/icse/ThompsonSCFSB0L25,DBLP:conf/kbse/LuD024,DBLP:journals/corr/abs-2410-19940}. 
Though many efforts are devoted to improving LLM performance in these proving agents~\cite{DBLP:conf/icse/ThompsonSCFSB0L25,DBLP:conf/kbse/LuD024,DBLP:journals/corr/abs-2410-19940,kozyrev2026rocqstarleveragingsimilaritydrivenretrieval,paraskevopoulou2026machinegeneratedmachinecheckedproofsverified}, we argue that enhancing symbolic provers themselves remains equally important. First, as revealed in existing studies~\cite{DBLP:conf/icse/Chen00025}, symbolic provers complement LLMs effectively.
If the improvement falls outside the proving capabilities of LLMs, the overall performance of the proving agent improves. 
Second, even if the improvement of symbolic provers falls within the capabilities of LLMs, we can reduce the expensive cost of invoking LLMs by offloading proving tasks to symbolic provers. Third, in many real-world scenarios, LLM-based approaches cannot be used at all. 
For example, small companies cannot afford the cost of their own LLM servers, and cannot use public services due to confidentiality concerns. Therefore, this paper aims to improve the performance of symbolic provers for automated theorem proving.

A key observation is that existing symbolic provers rely on the existing decision procedures~\cite{DBLP:conf/tacas/MouraB08,DBLP:conf/tacas/BarbosaBBKLMMMN22}, and do not work well in domains that lack decision procedures. 
However, in many such domains, existing proofs have already been written by human developers, and these proofs embed reusable proof strategies that serve as practical alternatives to the missing decision procedures.
Based on this observation, our basic idea is to enable symbolic provers to learn from existing proofs: we mine the proof strategies from existing proofs, formalize them as reusable tactics, and equip symbolic provers with this mined knowledge to handle cases where they would otherwise fail.

% To implement this idea, we draw inspiration from recent work on mining symbolic rules from data using LLMs~\cite{DBLP:conf/pakdd/LuoJXLHP25,DBLP:conf/iclr/ShojaeeMGFR25}. These approaches mine various rules such as horn clauses or mathematical equations from datasets such as knowledge graphs or numeric data. The success of these approaches suggests that LLMs can effectively extract symbolic knowledge from data, but none of these approaches are directly applicable to mining tactics from existing formal proofs. 

We propose \approach{}, the first approach that enables symbolic provers to mine proof strategies from existing formal proofs, using LLMs as the bridge to mine and formalize proof knowledge into reusable tactics.
Note that \approach{} does not train or fine-tune any neural network model; rather, it uses LLMs as an mining tool to surface proof strategies that already exist in proofs and make them accessible to symbolic provers.
Concretely, the workflow of \approach{} is as follows: i) we collect a corpus of formal proofs from existing verification projects as the knowledge source; ii) the LLM analyzes each proof to understand the underlying reasoning process; iii) the LLM identifies generalizable proof strategies embedded in the proof; iv) the LLM formalizes each strategy as a symbolic tactic; v) the mined tactics are integrated into the symbolic prover, enabling it to apply this acquired knowledge for automated proving.

Key challenges in this approach are: i) How can we ensure that mined tactics are useful? Many tactics generated by LLMs may be syntactically incorrect, or they may be overfitted to the specific proof from which they were mined, failing to generalize to new problems. ii) How can we retrieve appropriate tactics for a given proof state? Traditional retrieval methods such as BM25 rely on textual similarity. Since tactics are a structured representation with limited textual content, textual similarity measures are insufficient for identifying relevant tactics effectively.

To address the first challenge, we design a testing process that ensures: i) the tactic can be correctly represented in the symbolic system; and ii) the tactic is generalizable by demonstrating applicability to other theorems outside the mining dataset.
To address the second challenge, we record the theorem from which each tactic was mined, and retrieve tactics by using BM25 to match the theorems that the tactices are mined from. By indexing tactics via their source theorems, we effectively enrich the text for retrieval, as theorem statements carry richer textual information than tactics themselves.

\paragraph{Evaluation}
To evaluate \approach{}, we collect 11,725 theorems from Rocq standard library as the dataset for tactic mining. We choose the standard library because it contains a substantial number of theorems on commonly used data structures such as natural numbers, lists, and real numbers. 
To evaluate generalization capability, we collect a validation set containing 6,462 theorems from \coqcorn{}~\cite{DBLP:conf/mkm/Cruz-FilipeGW04}, including common algebraic structures (e.g., setoids, monoids, groups). We randomly select 93,424 different positions across these theorems' proof scripts to assess how broadly the mined tactic applies under different proof contexts.
To evaluate the enhancement to the symbolic prover, we select large, real-world open-source Rocq projects focused on software verification: \coqart{}~\cite{bertot2013interactive}, \extlib{}~\cite{coqextlib}, \vfa{}~\cite{Appel:SF3}, and \compcert{}~\cite{DBLP:conf/itp/KrebbersLW14}, in total \theoremsum{} theorems.
% \yxmodify{These projects are selected to represent a broad spectrum of real-world software verification tasks: \coqart{} and \vfa{} cover well-established properties of data structures and algorithms; \extlib{} reflects practical, community-driven Rocq development that is widely reused across diverse Rocq projects; and \compcert{} is an industrial-scale formally verified C compiler with machine-checked proofs of semantic preservation across multiple compilation passes, memory models, and target architectures, testing whether learned tactics generalize to large, heterogeneous proof developments far beyond toy examples.}{}
We collect a testing benchmark of \theoremsum{} theorems from these four benchmarks, and evaluate whether our approach can improve the proof capability of CoqHammer~\cite{DBLP:journals/jar/CzajkaK18}, the state-of-the-art symbolic prover in Rocq on these projects.
Our results show that using the tactics mined by Deepseek-V3.2~\cite{deepseekai2025deepseekv32}, \approach{} enhances the number of theorems proved by CoqHammer directly by \finalres{}. 
% Our retrieval method improves directly retrieving tactics using the state-of-the-art embedding model on Rocq data by \finalretrieve{}.
When integrating the improved CoqHammer into the state-of-the-art proving agent, Claude Code~\cite{claudecode2025}, the overall number of theorems proved increases by \finalresagent{}, and the token consumption is reduced by \finalrescost{}. 

\paragraph{Contribution}
The contributions of this paper are as follows:
\begin{itemize}
    \item We propose the first approach that enables symbolic provers to mine reusable proof strategies from existing formal proofs through LLMs, allowing the prover to apply this acquired knowledge at proof time without runtime LLM interaction and supporting deployment in constrained environments.
    \item We design a testing methodology that ensures mined tactics are generalizable to new proof goals outside the mining dataset.
    % \item We design an indirect tactic retrieval method on Rocq that overcomes the lack of textual information in Ltac tactics. %, outperforming the state-of-the-art embedding model specifically trained on Rocq data.
    \item We implement and evaluate \approach{} on multiple benchmarks, demonstrating that it enhances the automation capability of symbolic provers.
\end{itemize}

\section{Background}
\label{sec:background}

\textbf{The Rocq Proof Assistant.}
Rocq~\cite{CoqRefMan8.20} (formerly known as Coq) is an interactive theorem prover based on the Calculus of Inductive Constructions (CIC)~\cite{paulin2015introduction}.
Users state theorems and construct machine-checked proofs interactively. At each step, the system maintains a \emph{proof goal}: the proposition that remains to be proved. When a proof begins, the goal is the theorem statement itself. Users enter \emph{tactics}, each of which transforms the current goal into zero or more subgoals.
Once all subgoals are discharged, the Rocq kernel type-checks the proof term, guaranteeing correctness independently of the tactics that produced it.
Rocq provides a rich set of built-in tactics for common proof steps, including introducing hypotheses (\texttt{intros}), applying known lemmas (\texttt{apply}), rewriting with equalities (\texttt{rewrite}), and performing structural induction (\texttt{induction}).

\textbf{Ltac.}
Ltac~\cite{DBLP:conf/lpar/Delahaye00} is the tactic scripting language built into Rocq. It allows users to compose built-in tactics into higher-level, reusable proof procedures.
An Ltac definition binds a name to a tactic expression. The language supports pattern matching on the proof goal and hypotheses via the \texttt{match goal} construct, recursion, backtracking with the \texttt{+} and \texttt{||} operators, and looping via \texttt{repeat} and \texttt{do}. 
Ltac is widely used to encode domain-specific proof automation in Rocq projects. In this paper, \approach{} produces Ltac tactics as its reasoning strategy representation, so that the mined proof strategies can be directly loaded and invoked by the Rocq system.

\textbf{CoqHammer.}
CoqHammer~\cite{DBLP:journals/jar/CzajkaK18} is an automated theorem proving tool for Rocq. It adapts the "hammer" methodology that has been effective in other proof assistants (e.g., Sledgehammer for Isabelle~\cite{DBLP:conf/cade/BohmeN10, DBLP:conf/tphol/WenzelPN08}). The tool operates in three phases:
Given a proof goal, CoqHammer performs \emph{premise selection}, \emph{translation} to first-order logic with external ATP invocation, and \emph{proof reconstruction} inside Rocq to produce a fully checked proof term.
CoqHammer is the state-of-the-art general-purpose symbolic prover for Rocq~\cite{DBLP:journals/jar/CzajkaK18}. We use it as the baseline symbolic prover in our evaluation.
Despite its strength, CoqHammer does not learn new proof strategies from existing proofs; it relies on the fixed set of ATPs and reconstruction tactics. \approach{} addresses this limitation by mining reusable Ltac tactics from proof corpora and integrating them into CoqHammer's proof search.

\section{Overview}
\label{sec:overview}

In this section, we present an overview of \approach{}, a framework designed to mine reasoning strategies. We will use some examples to demonstrate the workflow of our approach.

\begin{figure*}[ht]
  \centering
  \includegraphics[width=0.6\linewidth]{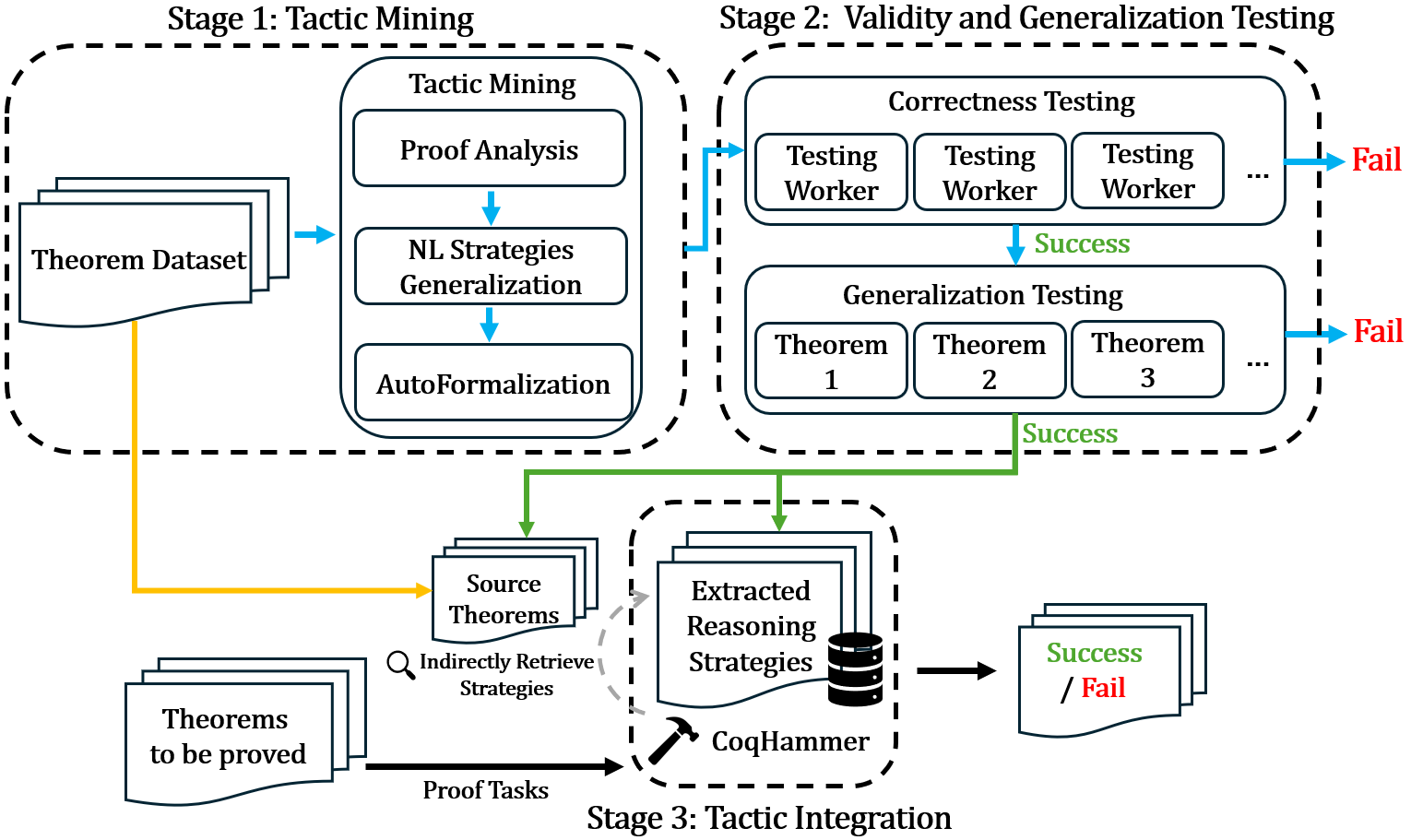}
  \caption{The workflow of \approach{}}
  \label{fig:overview}
\end{figure*}

\begin{figure}[t]
\begin{lstlisting}[language=Coq, basicstyle=\small\ttfamily,xleftmargin=0em]
(* Infinite trees with internal nodes labeled *)
CoInductive LTree (A:Type) : Type :=
  LLeaf : LTree A
| LBin : A -> LTree A -> LTree A -> LTree A.

(* An extensional equality on (LTree A) *)
CoInductive LTree_bisimilar :  LTree A -> LTree A -> Prop :=
  LTree_bisimilar_leaf : LTree_bisimilar LLeaf LLeaf
| LTree_bisimilar_bin : 
    forall (a:A) (t1 t'1 t2 t'2 : LTree A),
      LTree_bisimilar t1 t'1 ->
      LTree_bisimilar t2 t'2 ->
      LTree_bisimilar (LBin a t1 t2) (LBin a t'1 t'2).

Theorem LTree_bisimilar_label : 
  forall (p:path) (t t': LTree A),
    LTree_bisimilar t t' 
      -> LTree_label t p = LTree_label t' p.
\end{lstlisting}
\caption{Theorem to be proved in benchmarks}
\label{fig:running-ex}
\end{figure}

We consider the theorem \texttt{LTree\_bisimilar\_label} in~\Cref{fig:running-ex}, which comes from the \coqart{} project. This theorem describes a fundamental property of infinite trees. \texttt{LTree} is a tree data structure defined by \texttt{CoInductive}~\cite{DBLP:journals/lmcs/Capretta05}. Unlike inductive types that construct finite structures, coinductive types allow the definition of potentially infinite structures. The \texttt{CoInductive} keyword enables the construction of infinite trees where branches can continue indefinitely. The \texttt{LTree\_bisimilar} defines a property between infinite trees, establishing when two trees should be considered equivalent. This relation states that two trees are bisimilar if they have the same structure: either both are leaves, or both are binary nodes with the same label and bisimilar left and right subtrees. 

Here, a \texttt{path} is defined as a list of directions (left or right) that navigates through the tree structure. The function \texttt{L\_subtree} traverses the tree following the given path to extract a specific subtree, while \texttt{LTree\_label} extracts the label at the root of that subtree.
%and the definition is shown as following:
% \begin{lstlisting}[language=Coq, basicstyle=\ttfamily,xleftmargin=0em]
% Definition LTree_label {A:Type} (t:LTree A) (p:path) : 
%   option A :=
%   match L_subtree p t with
%   | None => None
%   | Some t' =>
%       match t' with
%       | LLeaf => None 
%       | LBin x _ _ => Some x 
%       end
%   end.
% \end{lstlisting}

The theorem \texttt{LTree\_bisimilar\_label} states that if two trees are bisimilar, they must have the same label at any given path. That establishes that bisimilarity preserves labels throughout the entire tree structure, which is a fundamental property for reasoning about infinite trees. 
However, this theorem cannot be directly proven by the automatic theorem prover CoqHammer.

The workflow of \approach{} is shown in~\Cref{fig:overview}. The entire framework is divided into three stages, where the first two steps are performed offline, similar to the training phase of machine learning, and the last step is performed online when proving new theorems.
\begin{itemize}
    \item \textbf{Stage 1: Tactic mining}. In this stage, we use large language models to mine reasoning strategies from a set of theorems and their proofs. 
    \item \textbf{Stage 2: Validity and generalization testing}. After mining the tactics, we need to ensure their validity and generalizability. We design specific tests to verify that the mined tactics can be applied in various scenarios.
    \item \textbf{Stage 3: Tactic integration}. We combine the mined tactics with symbolic provers to form a new prover with better theorem-proving capability. The prover uses these tactics to convert complex theorems into simpler goals until they can be automatically solved.
\end{itemize}

\subsection{Tactic Mining}

\paragraph{Proof analysis}
For each theorem and proof in the dataset, we ask the LLM to explain the purpose of each tactic in the proof, and what it means to use this tactic at the current position in natural language (NL). We analyze each tactic individually rather than asking the LLM to describe the proof as a whole, because the LLM tends to omit tactics whose effects are less apparent when summarizing an entire proof at once. Per-tactic analysis yields more detailed descriptions that capture the specific effect of every proof step. For example, the following is a theorem from the dataset: 
\begin{lstlisting}[language=Coq, basicstyle=\ttfamily]
Lemma append_neutral_r : 
  forall (i : positive), append i xH = i.
Proof. induction i; simpl; congruence. Qed.
\end{lstlisting}
where \texttt{positive} is an inductive data type that represents strictly positive integers in a binary way.

The LLM provides the following natural language description for the effect of each tactic:
\begin{lstlisting}[frame=single,breaklines=true,breakindent=0pt,basicstyle=\ttfamily,escapechar=@,basicstyle=\scriptsize\ttfamily,xleftmargin=0em]
### Step-by-step explanation
1. **induction i**
    - **Effect:** Performs induction on the binary positive integer i.
    So the induction generates three subgoals:
     - Base case: i = xH
     - Inductive case for xO p with IH for p ...
2. **simpl**
    - **Effect:** ...
    ...
\end{lstlisting}
We can observe that the LLM analyzed the roles of three tactics (\texttt{induction}, \texttt{simpl}, \texttt{congruence}) in this proof and the impact on the goals when each tactic is applied. Then, by providing the proof process and the meaning of each proof step as input, we guide the LLM to give a general reasoning strategy. 

\paragraph{NL strategies generalization}
In the second step, we require the LLM to summarize the given proof analysis into generalizable strategies\yxmodifyok{ based on the reasoning strategies learned during the LLM training phase}{}. After providing the above proof analysis to the LLM, the LLM gave the following general proof method:
\begin{lstlisting}[frame=single,breaklines=true,breakindent=0pt,basicstyle=\ttfamily,escapechar=@,basicstyle=\scriptsize\ttfamily,xleftmargin=0em]
## General Proof Strategies
### 1. **@\textcolor{red}{Structural Induction with Recursive Data Types}@**
This pattern is widely applicable to theorems about recursive data structures like lists, trees, and natural numbers, where properties can be proven by following the structure's definition.
...
### 2. **@\textcolor{red}{Definition-Driven Simplification Followed by Automated Equality}@
         @\textcolor{red}{Resolution}@**
This two-step approach combines:
1. **simpl tactic**: Expands and reduces terms based on their definitions in the context
2. **congruence tactic**: Automatically solves goals that become simple equalities after simplification
...
### 3. **@\textcolor{red}{Combination Pattern Recognition}@**
The proof demonstrates a common pattern: when a property is definitionally true for a recursive type, the proof often follows the template:
induction <variable>; simpl; <auto-solver like congruence/auto>
...
\end{lstlisting}
% \yxcomment{Pls solve the inconsistency in this example}

We can observe that the LLM provides general proof strategies. The first strategy is a standard one for recursive data types, which suggests the use of structural induction. The LLM also gives potential usage scenarios such as lists, trees, and natural numbers. The second strategy is to use automated equality resolution after simplification operations. 
The third strategy is a combination pattern that prescribes how to compose multiple tactics into a single reusable sequence, it describes how the three tactics can be combined to form higher-level proof methods.
We can observe that all strategies generalize beyond the current lemma and can be used in proving other theorems.

\paragraph{Autoformalization}
We require the LLM to formalize the new general proof strategies described in natural language into the tactic language used by the symbolic system. 
Since the theorems we selected are obtained from Rocq projects, for consistency, we formalize the new proof strategies into tactics represented in Ltac in Rocq~\cite{DBLP:conf/lpar/Delahaye00}.
Note that the role of autoformalization here is solely to translate natural language strategies into a symbolic representation; the correctness and generalizability of the resulting tactics are not assumed but are instead verified by the subsequent testing stage.
In this example, multiple tactics are formalized, and we show one tactic formalized from Strategy 3 as an example.
%Due to the previously combination strategy (Strategy 3), the LLM formalizes strategies by composing Strategy 1 and Strategy 2 into a single Ltac tactic, shown as the following:
\begin{figure}
\begin{lstlisting}[language=Coq, basicstyle=\small\ttfamily,xleftmargin=0em]
Ltac  structural_induction_recursive :=
  match goal with
  | [ |- forall p : positive, _ ] => induction p; simpl; try congruence
  | [ |- forall n : nat, _ ] => induction n; simpl; try congruence
  | [ |- forall l : list ?A, _ ] => induction l; simpl; try congruence
  | [ |- forall t : ?T, _ ] =>
      match T with
      | ?T' => induction t; simpl; try congruence
      end
  end.
\end{lstlisting}
\caption{One mined tactic from lemma \texttt{append\_neutral\_r} by \approach{}}
\label{fig:ex-tactic}
\end{figure}

\begin{figure*}[ht]
    \centering
    \includegraphics[width=1\linewidth]{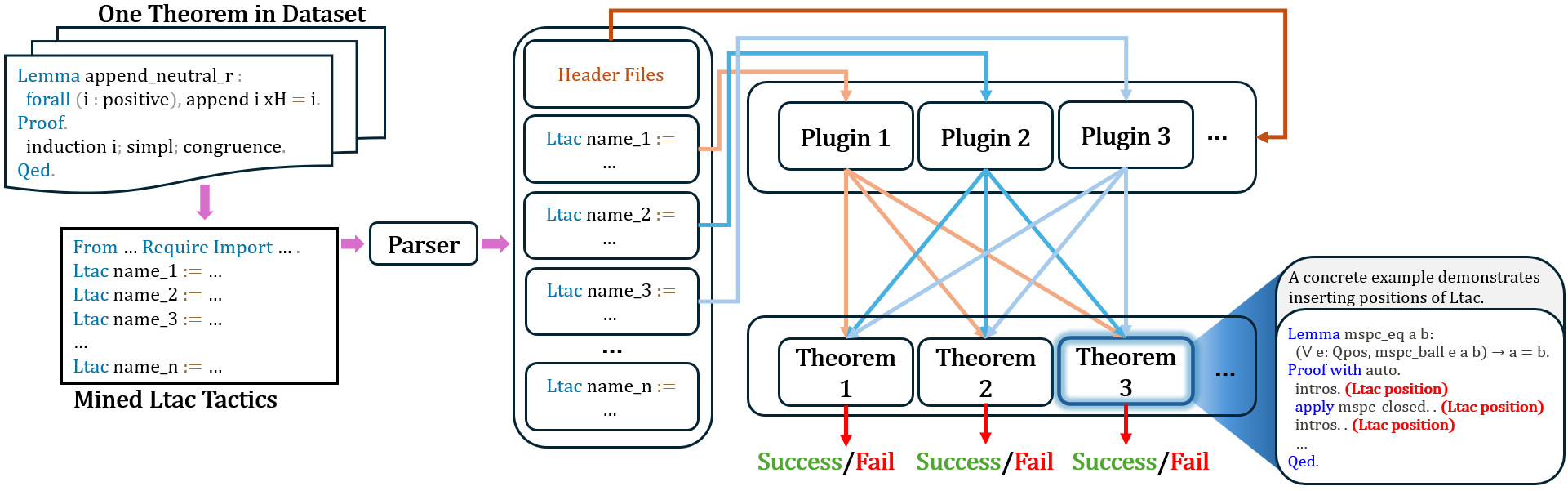}
    \caption{Workflow of generalization testing}
    \label{fig:generaliztion_testing}
\end{figure*}

\subsection{Validity and Generalization Testing}
Due to LLM hallucinations, the generated code by LLM may not be valid, so we need to test whether it is formalized correctly in Rocq Ltac. In addition, \approach{} also tests the generalization of tactics to ensure they can be used for proving other theorems. Note that a valid tactic is correct by construction in the Rocq system, and would never result in a incorrect proof.

To test validity, we employ a test-and-repair framework using the Rocq compiler. If compilation fails, we ask the LLM to repair the tactic. Tactics that fail after a fixed number of repairing attempts are discarded. To test generalizability, we use a validation set containing theorems that are not included in the mining dataset. We then apply the tactic into each position in the theorems' proving processes. If the proof statement changes after the application, we consider the tactic to have generalization capability and retain it; otherwise, we discard it.
For the retained tactics, \approach{} records the source theorem from which the tactic was mined, which will be used for retrieval when proving new theorems.

\subsection{Tactic Integration}
In this stage, \approach{} combines the mined tactics with a symbolic prover. Given a proof goal that cannot be directly resolved by the symbolic prover, we retrieve relevant tactics. As mentioned before, tactics contain limited textual information and direct retrieval of tactics is ineffective. Our approach uses the theorems for mining tactics as the bridge for retrieval. We first use BM25 to identify the theorems that are most similar to the current proof goal from the dataset for tactic mining, and return the tactics mined from the theorems. Next, we apply each of the retrieved tactics to transform the current proof goal and then apply the symbolic prover again. We construct an And-Or tree~\cite{DBLP:journals/cpc/ChauvinFGG04} to try different combinations of the above process until the theorem is proved.

%We design a proof algorithm (detailed in~\Cref{sec:tactic_integration}) that combines the learned tactics with the automatic theorem provers.
%Given a theorem to be proved, the algorithm applies a retrieval algorithm that, based on the current proof goal, identifies the most similar theorem from those recorded during generalization testing, and gets the corresponding tactics to apply.

For the theorem in~\Cref{fig:running-ex}, though CoqHammer cannot directly prove the theorem, the enhanced prover can retrieve the theorem \texttt{append\_neutral\_r} and use tactic \texttt{structural\_induction\_recursive} to convert the theorem into new goals that can be automatically solved by CoqHammer.

% The definition of \texttt{path} in the theorem is shown in the following:
% \begin{figure} 
% \begin{lstlisting}[language=Coq, basicstyle=\ttfamily,xleftmargin=0em]
% Inductive direction : Type :=
%   | d0 : direction (* left *)
%   | d1 : direction (* right *).

% Definition path := list direction.
% \end{lstlisting}
% \caption{Theorem to be proved in benchmarks}
% \label{fig:running-ex}
% \end{figure}

\section{Approach}
\label{sec:approach}
In this section, we present the detailed design of \approach{}. As illustrated in~\Cref{fig:overview}, the workflow of \approach{} consists of three main stages: i) \textbf{Tactic mining}. In this stage, \approach{} mines potential tactics from LLMs. ii) \textbf{Validity and generalization testing}. The testing checks the tactics that are syntactically valid and applicable to other theorems. iii) \textbf{Tactic integration}. \approach{} incorporates the mined tactics into the symbolic prover, employing a new retrieval method to select appropriate tactics based on the proof goal.

\subsection{Tactic Mining}
\label{sec:sub_learning}
% The purpose of tactic extraction is to extract generalizable tactics from LLMs. 
By inputting a dataset of theorems, we aim to mine the reasoning patterns that are already formed within proofs. This mining process consists of three steps: proof analysis, natural language strategy generalization, and autoformalization.

\begin{figure*}[t]
    \centering
    \hfill
    \begin{minipage}[t]{0.4\linewidth}
        \centering
        \includegraphics[width=\linewidth]{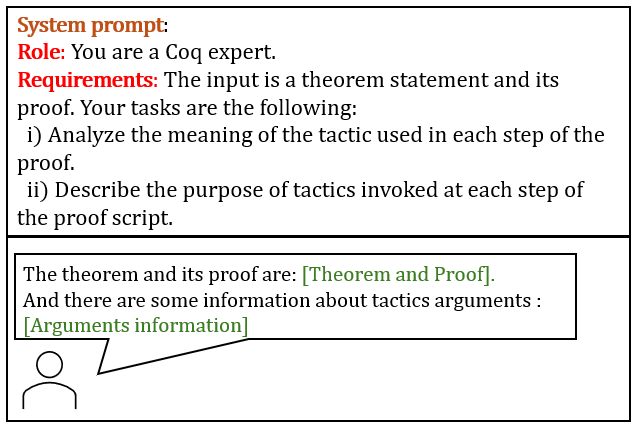}
        \caption{Prompts in proof analysis}
        \label{fig:proof_analysis}
    \end{minipage}
    \hspace{0.01\linewidth}
    \begin{minipage}[t]{0.4\linewidth}
        \centering
        \includegraphics[width=\linewidth]{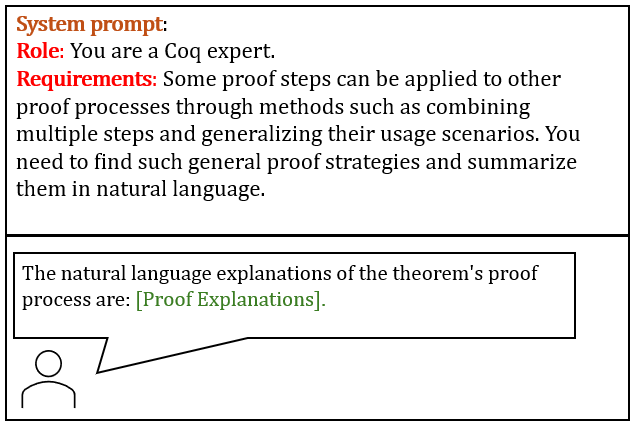}
        \caption{Prompts in natural language generalization}
        \label{fig:nl_generalization}
    \end{minipage}
    \hfill\hspace{0.0\linewidth}
\end{figure*}

\textbf{Proof Analysis.}
\label{sec:approach_1}
In the proof analysis step, we prompt LLMs to analyse the proof steps of given theorems. We require the LLM to perform two tasks. The prompts are shown in~\Cref{fig:proof_analysis}.
Specifically, the LLM performs two tasks per proof step: (1) explaining the semantic meaning of each tactic (e.g., what \texttt{rewrite} or \texttt{induction} does), and (2) describing the effect of each tactic on the current proof goal. To support accurate analysis, we supply the LLM with the types and definitions of inductive data referenced in the tactic parameters.

\textbf{Natural language strategies generalization.} 
In this step, we prompt the LLM to identify which steps from the previous proof analysis can be abstracted to a higher level. The prompt is shown in~\Cref{fig:nl_generalization}.
We use natural language descriptions for proof analysis and generalization, as LLMs are more powerful in understanding and generating natural language compared to symbolic language~\cite{wang2024speakoutsolvingsymbolrelated}. 

\begin{figure}
    \centering
    \includegraphics[width=0.85\linewidth]{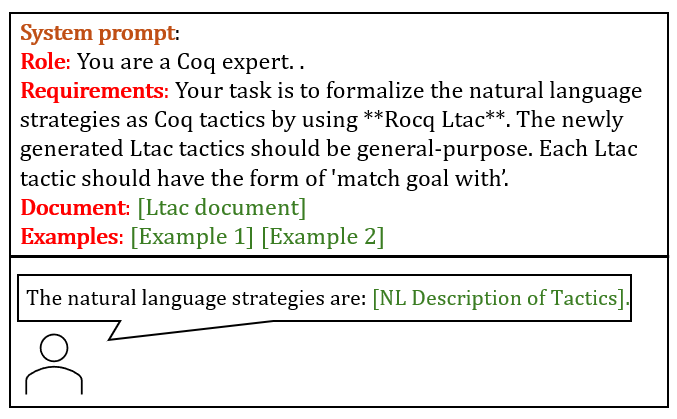}
    \caption{Prompts in Autoformalization}
    \label{fig:generalization}
\end{figure}

\textbf{Autoformalization.}
In the autoformalization step~\cite{wu2022autoformalizationlargelanguagemodels}, we use LLMs to convert the natural language descriptions of reasoning strategies into the formal language. The prompts used in the LLM are shown in~\Cref{fig:generalization}.
During autoformalization, the LLM translates the high-level strategic descriptions 
into Ltac~\cite{DBLP:conf/lpar/Delahaye00}, the tactic language of Rocq. 
We provide the LLM with Ltac documentation and examples to facilitate accurate 
translation, and instruct it to generate parameter-free tactics that match 
patterns in the proof goal.

\subsection{Validity and Generalization Testing}
\label{sec:sub_testing}

\textbf{Validity Testing.}
The purpose of this stage is to test the tactics mined from the LLM.
In validity testing, we design a test-and-repair framework to test whether the mined tactics are valid.
We use the symbolic verifier to compile each tactic. If the verifier finishes compilation without errors, we consider the tactic to be valid within the logical system.
If the verifier reports errors, we provide both the error message and the current tactic back to the LLM, prompting it to repair the errors. If the number of repair attempts exceeds a threshold, we discard this tactic.

\textbf{Generalization Testing.}
Generalization testing evaluates whether the tactics can be applied to theorems in other projects. We collect a large number of theorems from other projects as the benchmark. 
The workflow is shown in~\Cref{fig:generaliztion_testing}.
The overall workflow proceeds as follows: first, since a single theorem may yield multiple tactics, we employ a parser to decompose the tactic into individual tactic definitions along with their required header files and type definitions. This decomposition step is crucial as it isolates each tactic as an independent unit while preserving all necessary dependencies for compilation.
Then, for each mined tactic, \approach{} automatically generates a Rocq plugin that wraps the tactic with all its required dependencies and exposes it through a named interface. This plugin-based method allows each proof script to load a tactic via a single command, avoiding repeated script modifications and dependency conflicts across different tactics.

For generalization testing, we insert each mined tactic after every original tactic in the benchmark proof scripts (except for the last tactic that closes the goal), and check whether it successfully produces a new proof state. As shown in Figure~\ref{fig:generaliztion_testing} theorem 3, a concrete example demonstrates the inserting positions of the tactics (marked by "Ltac position").
A tactic is considered to be generalizable if, given the current proof state and theorem, it produces new proof states without reporting errors (a tactic without errors but makes no changes to the proof state is discarded).
We record the number of successful applications across all benchmark positions as a metric of generalization capability. Specifically, we consider a tactic to have generalization capability if its application success rate across all benchmark positions falls within $[10\%, 100\%)$ (details in \Cref{sec:rq1}). A tactic with a success rate below $10\%$ is considered too narrow to use reliably. 
Conversely, a tactic that succeeds in $100\%$ of positions is considered over-general: such a tactic typically only introduces a fresh variable to alias an existing term in the proof context, which changes the proof state without making meaningful reasoning progress.

\subsection{Tactic Integration}
\label{sec:tactic_integration}
Tactics that pass the generalization testing are combined with a symbolic prover for better theorem proving. 
%are integrated into the tactic library of the symbolic prover to obtain an enhanced symbolic prover. 
%However, existing automated theorem provers do not support the integration of user-defined tactics, making it difficult to directly leverage the newly mined tactics within the prover pipeline.
Since existing symbolic provers do not support the utilization of the tactics presented in the code, we design an algorithm to alternately invoke the symbolic prover and apply tactics, shown as algorithm \textsc{Solve} in \Cref{alg:proof_search}. %that combines the mined tactics with the automated theorem prover, enabling them to work in a unified solving process.
%\approach{} introduces the algorithm \textsc{Solve}, as shown in~\Cref{alg:proof_search}. 
The algorithm models proof search by constructing an And-Or tree~\cite{DBLP:journals/cpc/ChauvinFGG04}. Each proof goal forms an \textsc{Or}-node: it is solved if \emph{any} tactic succeeds. Each tactic application forms an \textsc{And}-node: it requires \emph{all} resulting subgoals to be proved.

\begin{algorithm}[ht]
\caption{And-Or Tree Proof Search with Tactic Retrieval}
\label{alg:proof_search}
\begin{algorithmic}[1]
\Function{Solve}{$goal$}
    \State $root \gets \Call{OrNode}{goal}$
    \State $queue \gets \Call{Queue}{\,}$; \ $queue.\Call{Enqueue}{root}$
    \While{$queue$ is not empty $\land$ $\lnot\, root.solved$}
        \State $n \gets queue.\Call{Dequeue}{\,}$
        \If{\Call{ApplyATP}{$n.goal$}}
            \State \Call{Propagate}{$n$}; \ \textbf{continue}
        \EndIf
        \State $\mathit{Thms} \gets \Call{RetrieveTheorems}{n.goal,\ k}$ 
        \State $T \gets \Call{GetTactics}{\mathit{Thms}}$ 
        \For{\textbf{each} $t \in T$}
            \State $subgoals \gets \Call{Apply}{t,\ n.goal}$
            \State $a \gets \Call{AndNode}{\,}$; \ $n.\Call{AddChild}{a}$
            \If{$subgoals = \emptyset$}
                \State \Call{Propagate}{$a$}
            \Else
                \For{\textbf{each} $g' \in subgoals$}
                    \State $c \gets \Call{OrNode}{g'}$; \ $a.\Call{AddChild}{c}$
                    \State $queue.\Call{Enqueue}{c}$
                \EndFor
            \EndIf
        \EndFor
    \EndWhile
    \State \Return $root.solved$
\EndFunction
\end{algorithmic}
\end{algorithm}

\begin{table}[ht]
    \centering
    \caption{Impact of $k$ value on \Cref{alg:proof_search}}
    \label{tab:k_value}
    \begin{tabular}{ccc}
      \toprule
      \textbf{$k$ value} & \textbf{Proved} & \textbf{Time Consumption}  \\
      \midrule
      10 &  77  &  2212.89s \\
      20 & 85   & 3571.96s           \\
      30 & 87   & 4103.70s  \\
      \bottomrule
    \end{tabular}
  \end{table}
Given a proof goal, \textsc{Solve} first attempts to close it with the automated theorem prover. If that fails, it retrieves the top-$k$ most similar theorems using \textsc{RetrieveTheorems}.
Since Ltac is a structured representation, term-frequency algorithms such as BM25~\cite{DBLP:journals/ftir/RobertsonZ09} cannot be directly applied. 
To address this, \approach{} records the source theorem from which each tactic was 
mined during generalization testing. In our algorithm, the current proof goal is 
compared against these source theorems using a retrieval algorithm. 
Then, it locates the tactics mined from these theorems using the \textsc{GetTactics} 
function.
Each tactic application creates an \textsc{And}-node whose children are new \textsc{Or}-nodes for the resulting subgoals, which are enqueued for subsequent processing. When a node is solved, \textsc{Propagate} walks upward through the tree: an \textsc{Or}-node is marked solved when any child succeeds; an \textsc{And}-node is marked solved only when all its children are solved.

The parameter $k$ controls how many similar theorems are retrieved at each node. 
A larger $k$ increases the chance of finding a useful tactic but also expands the search space and increases time cost. 
We randomly select 50 samples from each benchmark (mentioned in \Cref{sec:intro}), for a total of 200 samples.
we set $k$ to 10, 20, and 30, respectively, and record the number of proved theorems and the time consumed for each setting. and the result is shown in~\Cref{tab:k_value}.
We evaluate this trade-off and set $k=20$ for our approach.

\section{Evaluation}
We focus on the following research questions:
\begin{itemize}
    \item \textbf{RQ1:} How effective is \approach{} in mining tactics?
    \item \textbf{RQ2:} To what extent do the tactics mined from LLMs by \approach{} improve the success rate of symbolic provers?
    \item \textbf{RQ3:} Do the mined tactics benefit LLM agents?
    \item \textbf{RQ4:} How does \approach{} perform compared to existing lemma extraction methods?

    \item \textbf{RQ5:} How do mining methods affect the mined tactics?
    \item \textbf{RQ6:} To what extent does the generalization testing in \approach{} impact the improvement of symbolic provers?
  \end{itemize}

\textbf{Implementation.}
We have implemented \approach{} on Rocq 8.20.0. To obtain the argument information in~\Cref{sec:sub_learning}, we modified parts of the Rocq compiler to extract argument types and type definitions. We selected CoqHammer, the state-of-the-art automated theorem prover in Rocq~\cite{DBLP:journals/jar/CzajkaK18}, as the symbolic prover for evaluating performance improvements. 

\textbf{Dataset.} 
We selected different benchmarks from Rocq open-source projects for different stages in \approach{}. We collected 11,725 theorems from the Rocq standard library for mining tactics. We collected 6,462 theorems from the \coqcorn{}~\cite{DBLP:conf/mkm/Cruz-FilipeGW04} dataset, which contains theorems on common algebraic structures (e.g., setoids, monoids, groups) and large proof procedures for complex properties, for testing generalization. For each candidate tactic, we insert it at all 93,424 different positions across these 6,462 theorems' proof processes to assess how broadly the tactic applies under different proof contexts.
To evaluate the improvement to symbolic provers, we selected the following four benchmarks to evaluate different verification tasks:
\begin{itemize}
    \item \coqart{}~\cite{bertot2013interactive}: a Rocq textbook covering well-established properties of data structures and algorithms.
    \item \extlib{}~\cite{coqextlib}: a community-driven Rocq library that is widely reused across diverse Rocq projects.
    \item \vfa{}~\cite{Appel:SF3}: the third volume of Software Foundations covering well-established properties of verified functional algorithms.
    \item \compcert{}~\cite{DBLP:conf/itp/KrebbersLW14}: an industrial-scale formally verified C compiler with machine-checked proofs of semantic preservation across multiple compilation passes, memory models, and target architectures.
\end{itemize}
Together, they span varying levels of proof complexity and application domains, making them a rigorous testbed for evaluating the generalization of \approach{}.

\textbf{Configurations.}
We used OpenAI GPT-5.2~\cite{openaigpt52} and Deepseek-V3.2~\cite{deepseekai2025deepseekv32} for evaluation experiments. For the tactic repair process in~\Cref{sec:sub_testing}, we set the threshold to 3. 
The total API cost for tactic mining was \$1,528.65, of which Deepseek-V3.2 accounted for \$108.93 and GPT-5.2 for \$1,419.72, well below the cost of GPU infrastructure to host comparable models locally.
We set CoqHammer's timeout to default value, which is 20 seconds, and for each theorem to be proved, we set the timeout to 600 seconds.
% For $k$ value in \Cref{alg:proof_search}, we researched the impact of $k$ on the performance of \approach{} in RQ2.
We select mined tactics whose application success rate falls within $[10\%, 100\%)$ (at least $10\%$ but less than $100\%$) in generalization testing.
% For $k$ value in \Cref{alg:proof_search}, we set it to 20 by default (see \Cref{sec:rq2}). 
For the retrieval algorithm, since our approach targets resource-constrained settings, embedding-based models that require high-performance GPUs are not applicable for tactic retrieval. We therefore choose BM25 to index tactics, which runs efficiently on CPUs~\cite{DBLP:journals/ftir/RobertsonZ09}. Prior work has also shown that BM25 outperforms embedding-based models for retrieval in theorem proving~\cite{DBLP:conf/icse/ThompsonSCFSB0L25}.
All experiments were conducted on a server equipped with four 26-core Intel Xeon Platinum 8270 CPUs, 256 GB RAM, and running Ubuntu 24.04.4. 

\subsection{\textbf{RQ1:} How effective is \approach{} in mining tactics?}
\label{sec:rq1}
\begin{figure}
    \centering
    \includegraphics[width=1\linewidth]{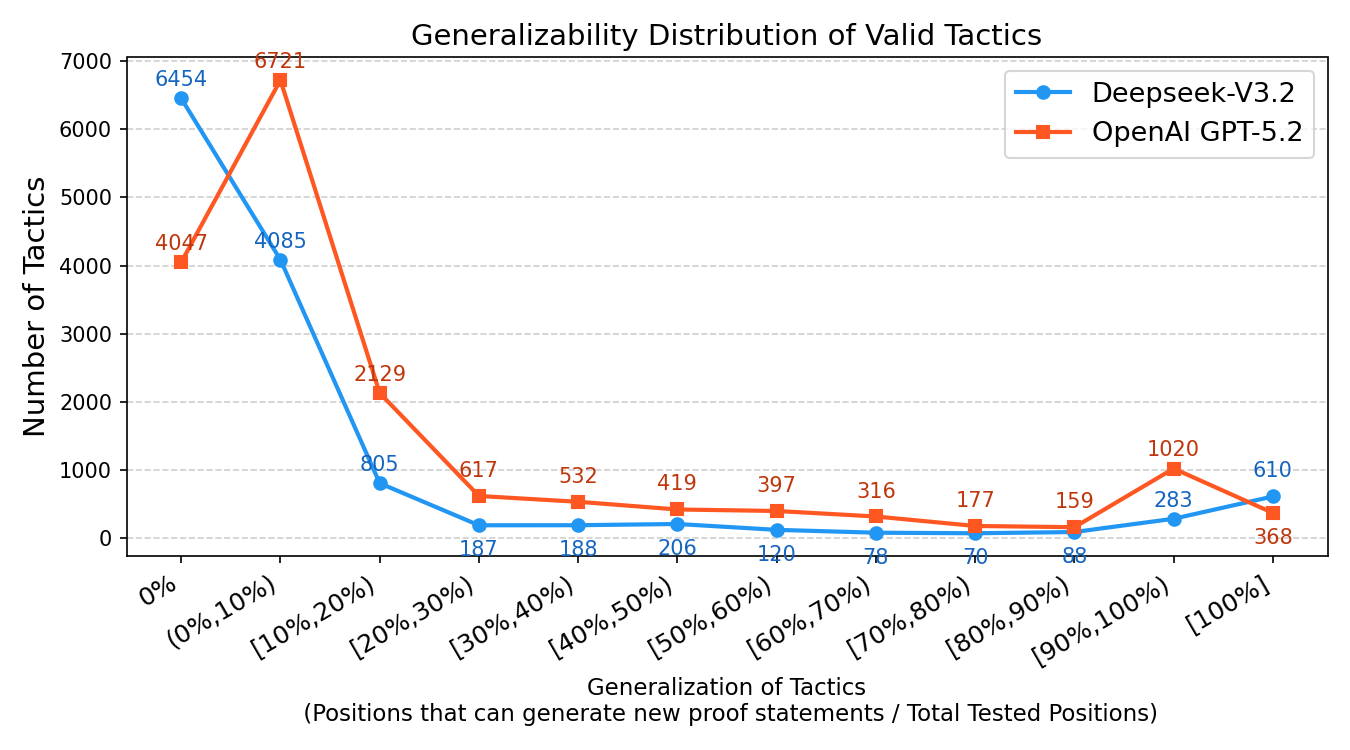} 
    \caption{Distribution of Tactic Generalizability} 
    \label{fig:tactic_distribution}
\end{figure}

\paragraph{Procedure}
This experiment evaluates the tactic mining of \approach{}. We apply \approach{} to mine tactics from 11,725 theorems in the Rocq standard library. For validity testing, we record the number of tactics obtained by OpenAI GPT-5.2 and Deepseek-V3.2 at each stage of the \approach{}. For generalization testing, we count the generalizability distribution of valid tactics, and classify all valid tactics into groups based on their success rate with an interval of 10\%, as shown in~\Cref{fig:tactic_distribution}.

\begin{table}[ht]
\centering
\caption{Detailed Statistics for Tactic Mining}
\label{tab:rq1}
\begin{tabular}{lcc} 
\toprule
\multicolumn{1}{r}{}      & \multicolumn{1}{l}{\textit{OpenAI GPT-5.2}} & \multicolumn{1}{r}{\textit{Deepseek-V3.2}}  \\ 
\midrule
Mined Tactics        & 49909                               & 35408                               \\ 
\midrule
Valid Tactics    & 16902                             & 13174                               \\ 
Generalizable Tactics ([10\%,100)) & 5766                                & 2025                                \\
\bottomrule
\end{tabular}
\end{table}

\paragraph{Results}
As shown in~\Cref{tab:rq1}, ``mined tactics'' represents the number of natural language tactics obtained from LLMs, ``valid tactics'' represents the number of tactics that pass the Rocq compiler, and ``generalizable tactics'' represents the number of tactics whose generalization testing success rate falls within $[10\%, 100\%)$.

We observe that OpenAI GPT-5.2 mines more tactics (49,909) compared to Deepseek-V3.2 (35,408). After validity testing, the pass rates are comparable: 33.87\% for OpenAI GPT-5.2 and 37.21\% for Deepseek-V3.2. 
For generalization testing, we observe that LLMs still exhibit severe hallucinations in tactic mining:
most generated tactics are incorrect and lack generalization capability, even though LLMs believe they are correct and generalizable.
We further observe notable differences in the generalization distributions across models, and the majority of mined tactics fall below the $10\%$ success rate.
This also indicates the necessity of the testing process in \approach{}.

\paragraph{Case Study}
To illustrate the diversity of mined tactics, we present a representative example beyond the induction-based strategy shown in~\Cref{fig:ex-tactic}. The following tactic shows a tactic mined by \approach{} that operates in a fundamentally different way.
\begin{lstlisting}[language=Coq, basicstyle=\small\ttfamily,xleftmargin=0em]
Ltac apply_constructor_auto_217 :=
  match goal with
  | [ |- ?R ?A ?B ] =>
      try (apply permA_swap; auto);
      try (intuition; auto)
  end.
\end{lstlisting}

This tactic first pattern-matches the current proof goal against a relational structure \texttt{?R ?A ?B}, then attempts to apply a lemma \texttt{permA\_swap} and invokes built-in Rocq tactics \texttt{auto} and \texttt{intuition} to discharge remaining subgoals. Unlike the induction-based tactic in~\Cref{fig:ex-tactic}, this example shows that \approach{} can mine strategies that reason about the goal's structure, apply project-specific lemmas, and compose multiple existing tactics in Rocq.

\begin{rqbox}
\textbf{Answer to RQ1:} 
\approach{} is able to mine generalizable tactics from LLMs, and the testing process is necessary to filter out invalid or non-generalizable tactics.

%\approach{} demonstrates effectiveness in extracting tactics. Despite LLMs exhibiting severe hallucinations with most extracted tactics being incorrect or lacking generalization capability, our approach successfully filters useful tactics from tens of thousands of LLM-generated candidates.
\end{rqbox}

\subsection{\textbf{RQ2:} To what extent do the tactics mined from LLMs by \approach{} improve the success rate of symbolic provers?}
\paragraph{Procedure}
We use \compcert{}, \extlib{}, \coqart{}, and \vfa{} four benchmarks, \theoremsum{} theorems in total, to test the automatic capability of the prover. We compare the \approach{}-enhanced CoqHammer against the original CoqHammer. For each experiment, we use the default configuration of CoqHammer and the timeout is set to 20 seconds.

\begin{table*}[ht]
\centering
\caption{Performance of CoqHammer with and Tactics Mined By \approach{}}
\label{tab:rq2}
\begin{tabular}{cccr@{\hspace{2pt}}lcr@{\hspace{2pt}}l}
\toprule
 & &\multicolumn{3}{c}{\textit{OpenAI GPT-5.2}} & \multicolumn{3}{c}{\textit{Deepseek-V3.2}} \\
 \cmidrule(lr){3-5} \cmidrule(lr){6-8}
 & \textit{\#Lemma} &CoqHammer & \multicolumn{2}{c}{\begin{tabular}[c]{@{}c@{}}CoqHammer\\+\approach{}\end{tabular}} & CoqHammer & \multicolumn{2}{c}{\begin{tabular}[c]{@{}c@{}}CoqHammer\\+\approach{}\end{tabular}} \\ 
 \midrule
\compcert{} & 4840 & 1022 &  \textbf{1269} & ($\uparrow$ \textbf{24.17\%}) & 1022 & \textbf{1276} & ($\uparrow$ \textbf{24.85\%})   \\ 
\extlib{} & 187 & 79 & \textbf{100} & ($\uparrow$ \textbf{26.58\%}) & 79 & \textbf{99} & ($\uparrow$ \textbf{25.32\%}) \\ 
\coqart{} & 842 & 411 & \textbf{482} & ($\uparrow$ \textbf{17.27\%}) & 411 & \textbf{498} & ($\uparrow$ \textbf{21.17\%})   \\ 
\vfa{} & 330 & 80 &  \textbf{99} &  ($\uparrow$ \textbf{23.75\%}) & 80 &  \textbf{99} & ($\uparrow$ \textbf{23.75\%})   \\ 
\midrule
Total & \theoremsum{} & 1592 & \textbf{1950} & ($\uparrow$ \textbf{22.49\%}) & 1592 & \textbf{1972} & ($\uparrow$ \textbf{23.87\%})   \\ 
\bottomrule
% Total & 198.00 & \textbf{220.00} ($\uparrow$ \textbf{11.28\%)} \\ \hline
\end{tabular}
\end{table*}

\paragraph{Results}
As shown in~\Cref{tab:rq2}, integrating the tactics mined by \approach{} significantly improves the performance of CoqHammer across all four benchmarks. Using tactics mined by OpenAI GPT-5.2, CoqHammer proves 1950 theorems, compared to 1592 theorems by the original CoqHammer, representing a 22.49\% overall improvement. The improvements on individual benchmarks range from 17.27\% on \coqart{} to 26.58\% on \extlib{}. Using tactics mined by Deepseek-V3.2, CoqHammer proves 1972 theorems, achieving 23.87\% overall improvement. 
While both models demonstrate substantial improvements, the performance of tactics mined from different LLMs is different.
This difference indicates that the reasoning strategies mined from different LLMs have an impact on the performance of \approach{}.
Notably, as shown in~\Cref{tab:rq1}, GPT-5.2 produces nearly three times more generalizable tactics than Deepseek-V3.2 (5,766 vs.\ 2,025), yet its proving count is slightly lower (1,950 vs.\ 1,972). This suggests that a larger number of tactics does not necessarily lead to better proving performance. A larger tactic pool introduces more noise during retrieval, causing the prover to spend time on less effective tactics before finding the right one. 

\begin{rqbox}
\textbf{Answer to RQ2:} 
The tactics mined by \approach{} significantly enhance the capability of CoqHammer by 17.27\% to 26.58\%, and the improvements are related to the reasoning capability of the underlying LLMs, suggesting further improvements as LLMs advance.
%The tactics extracted by \approach{} significantly enhance the capability of symbolic provers. OpenAI GPT-5.2 achieves a 24.34\% overall improvement, while Deepseek-V3.2 achieves an 11.81\% improvement across all benchmarks. These results demonstrate that the extracted tactics effectively improve provers' capabilities.
\end{rqbox}

\subsection{\textbf{RQ3:} Do the mined tactics benefit LLM agents?}
\paragraph{Procedure}
Due to the high cost of LLM-based agents, we sample 50 theorems from the test set of each benchmark, in total 200 theorems, the same as in RQ2, to evaluate whether \approach{} can benefit the LLM-based agent. 
We select Claude Code~\cite{claudecode2025}, the state-of-the-art LLM-based agent, powered by Claude Sonnet 4.6~\cite{anthropic2026sonnet46} as the base model. ~\cite{paraskevopoulou2026machinegeneratedmachinecheckedproofsverified} shows that Claude Code also performs well on Rocq theorem proving.

We write three categories of skills to instruct Claude Code to write theorem proofs. The first category covers the base proving pipeline, guiding the agent to locate the target theorem within a project and verify whether a constructed proof is correct. 
This category is common to all three settings in the following evaluation.
The second category is CoqHammer-specific: it instructs the agent on how to import the required header files into the proof script and which tactics to apply. 
The third category is \approach{}-specific: it explains to the agent how to use the \approach{}-enhanced prover.
Both categories direct the agent to first attempt to prove the theorem with the respective tool, and only if that attempt fails, to synthesize a proof on its own.
For each task, we grant Claude Code three tool permissions: \texttt{bash}, \texttt{read}, and \texttt{edit}, allowing the agent to execute shell commands, read files, and edit files. We set a timeout of 600 seconds for each proof task.

We evaluate three settings: i) Claude Code proves theorems without access to CoqHammer; ii) Claude Code proves theorems with access to CoqHammer via the corresponding skills; iii) Claude Code proves theorems with the \approach{}-specific skills. For each setting, we record the number of theorems successfully proved and the total token consumption of Claude Code.

\begin{table}[ht]
\centering
	\caption{Performance of the LLM Agents}
	\label{tab:rq3}
\begin{tabular}{lccc}
\toprule
             & Claude Code & \multicolumn{1}{c}{\begin{tabular}[c]{@{}c@{}}Claude Code \\+ CoqHammer\end{tabular}} & \multicolumn{1}{c}{\begin{tabular}[c]{@{}c@{}}Claude Code \\+ \approach{}\end{tabular}}  \\
\midrule
\#Thm & 71 & 101 &  111 ($\uparrow 9.90\%$)  \\
\midrule
Tokens  & 59899928   & 38346691             &   34317809      ($\downarrow 10.51\%$)          \\
\bottomrule
\end{tabular}
\end{table}

\paragraph{Results}
As reported in~\Cref{tab:rq3}, the \approach{}-enhanced agent proved 111 out of 200 randomly sampled theorems, compared to 101 by the CoqHammer-enhanced agent (a 9.90\% improvement). This result demonstrates that \approach{} is more effective than CoqHammer when integrated into a LLM-based proving agent.
Meanwhile, the token consumption decreased from 38346691 to 34317809, a 10.51\% reduction. This demonstrates that \approach{} not only enhances the proving capability of LLM agents but also reduces their token cost.

\begin{rqbox}
\textbf{Answer to RQ3:} 
The mined tactics from \approach{} can benefit LLM agents by improving their theorem-proving success rate and reducing token consumption.
\end{rqbox}

\subsection{\textbf{RQ4:} How does \approach{} perform compared to existing lemma extraction methods?}
\paragraph{Procedure}
As will be discussed in \Cref{sec:related_work}, a recently posted work, \stratrocq{}~\cite{fang2025proofstrategyextractionllms}, mines lemmas from the LLM reasoning process to enhance symbolic provers.
To compare our approach with \stratrocq{}, we implemented their approach to extract lemmas from the LLM reasoning process over the same dataset used in RQ1. We then integrate the extracted lemmas into CoqHammer and evaluate the performance improvement on the same four benchmarks used in RQ2.

\begin{table}
\centering
\caption{Comparison with the lemma extraction approach \stratrocq{}}
\label{tab:rq4}
\begin{tabular}{cccc} 
\toprule
                             & CoqHammer & \begin{tabular}[c]{@{}c@{}}\quad CoqHammer\\+\stratrocq{}\end{tabular} & \begin{tabular}[c]{@{}c@{}}CoqHammer\\+\approach{}\end{tabular}  \\ 
\midrule
\compcert{} &  1022    & 1072 & 1276 \\ 
\midrule
\extlib{}   &  79       & 73 & 99 \\ 
\midrule
\coqart{}   &  411     & 379 &  498\\ 
\midrule
\vfa{}      &  80      & 71 & 99  \\
\midrule
Total & 1592  & 1595 & 1972  \\
\bottomrule
\end{tabular}
\end{table}

\paragraph{Results}
The results are shown in~\Cref{tab:rq4}.
By reproducing \stratrocq{}, we extracted 2,977 new lemmas and integrated them into CoqHammer's context.
In our experiments, we observe that, under the current experimental setup, \stratrocq{} does not effectively enhance the symbolic prover. This discrepancy from the existing evaluation results reported in the \stratrocq{} paper may be attributed to the differences in dataset selection. In their paper, the dataset for mining lemmas and the dataset for testing are more related, and may come from the same project, while in our evaluation the two sets are from strictly different projects, better matching real-world scenarios. This result suggests that the tactics mined by our approach are more generalizable across different projects compared to the lemmas mined by \stratrocq{}.

\begin{rqbox}
\textbf{Answer to RQ4:} 
\approach{} significantly outperforms \stratrocq{}, especially when the dataset for mining and the dataset for testing are less closely related.

\end{rqbox}

% \subsection{\textbf{RQ5:} How do extraction methods affect the performance of \approach{}?}
\subsection{\textbf{RQ5: } How do mining methods affect the mined tactics?}
\paragraph{Procedure}
To evaluate the contribution of different components in the mining process, we conduct an ablation study using Deepseek-V3.2. Specifically, we compare three configurations: (1) the original \approach{} with all components, (2) \approach{} without the natural language analysis step (the first step in \Cref{sec:approach_1}), and (3) \approach{} without the natural language generalization step (the second step in \Cref{sec:approach_1}). We apply these configurations to mine tactics from the same dataset and measure the number of mined tactics, valid tactics, and generalizable tactics.

\begin{table*}[ht]
\centering
\caption{The ablation study of the mining methods}
\label{tab:rq5}
\begin{tabular}{lccccc}
\toprule
    &  Original Tactics  &  \multicolumn{2}{c}{\begin{tabular}[c]{@{}c@{}}-NL \\Analysis\end{tabular}} &  \multicolumn{2}{c}{\begin{tabular}[c]{@{}c@{}}-NL \\Generalization\end{tabular}}  \\
\cmidrule(lr){3-4} \cmidrule(lr){5-6}
                            &  & \# & Change & \# & Change \\
\midrule
Mined Tactics        & 35408    & 23256     & \textbf{$\downarrow$ 34.31\%}     & 38003   & \textbf{$\uparrow$ 7.33\%}              \\
\midrule
Valid Tactics    & 13174    &  6371     & \textbf{$\downarrow$ 51.64\%}      &  4415   & \textbf{$\downarrow$ 66.49\%}              \\

Generalizable Tactics ([10\%, 100\%)) & 2025     &   576     & \textbf{$\downarrow$ 71.56\%}      &  231    & \textbf{$\downarrow$ 88.59\%}              \\
\bottomrule
\end{tabular}
\end{table*}

\paragraph{Results}
As shown in Table \ref{tab:rq5}, the ablation study reveals the importance of both natural language analysis and generalization steps in \approach{}. 
Removing the natural language analysis step leads to a 34.31\% reduction in mined tactics, with more drops in valid tactics (51.64\% decrease) and generalizable tactics (71.56\% decrease) test pass rates. This indicates that the analysis step is helpful for guiding LLMs to generate higher-quality tactic candidates.

Without the natural language generalization step, valid tactics drop by 66.49\%. This indicates that directly formalizing proof analysis is more challenging for LLMs than describing tactics in natural language. Generalizable tactics decreased by 88.59\%. This demonstrates that the tactic generalization step plays a crucial role in producing the final usable generalizable tactics.

\begin{rqbox}
\textbf{Answer to RQ5:} 
Both the proof analysis and tactic generalization steps are essential for \approach{} to mine tactics from LLMs. 
\end{rqbox}

\subsection{\textbf{RQ6:} To what extent does the generalization testing in \approach{} impact the improvement of symbolic provers?}
\paragraph{Procedure}
To evaluate the impact of generalization testing, we compare two settings using tactics mined from Deepseek-V3.2: (1) CoqHammer enhanced with all mined tactics, without applying generalization testing, and (2) CoqHammer enhanced only with tactics that have passed generalization testing (passing rate in [10\%, 100\%)). We use the same four benchmarks and experimental configuration as in RQ2 to measure the performance difference.

\begin{table}[ht]
\centering
\caption{Impact of Generalization Testing on Performance with \approach{}}
\label{tab:rq6}
\begin{tabular}{llll}
\toprule
& CoqHammer & \begin{tabular}[c]{@{}l@{}}\quad CoqHammer\\+\approach{}\end{tabular} & \begin{tabular}[c]{@{}l@{}}CoqHammer\\+\approach{}\\-(GT)\end{tabular}  \\
\midrule
\compcert{} &  1022     & 1276 & 1092 \\ 
\midrule
\extlib{}   &  79        & 99 &  79 \\ 
\midrule
\coqart{}   &  411        & 498 & 460 \\ 
\midrule
\vfa{}      &  80        & 99 & 77 \\
\midrule
Total & 1592  & 1972  & 1708 \textbf{($\downarrow$ \textbf{13.39\%})} \\
\bottomrule
\end{tabular}
\end{table}

\paragraph{Results}
As shown in~\Cref{tab:rq6}, we observe that for \compcert{} and \coqart{}, CoqHammer enhanced with tactics that have not passed generalization testing shows improvement over the original CoqHammer. However, their performance is lower than CoqHammer enhanced with tactics that have passed generalization testing. This demonstrates the effectiveness of generalization testing. Without generalization testing, many ineffective tactics are present, preventing the retrieval of effective tactics.

\begin{rqbox}
\textbf{Answer to RQ6:}
The results demonstrate that the generalization testing is effective in improving the performance of \approach{}.
\end{rqbox}

\section{Related Work}
\label{sec:related_work}
% Recent work has explored extracting structured knowledge from large language models. Luo et al.~\cite{DBLP:conf/pakdd/LuoJXLHP25} attempt to extract logic rules from knowledge graphs using LLMs. Shojaee et al.~\cite{DBLP:conf/iclr/ShojaeeMGFR25} use LLMs to discover mathematical equations from data. Zhang et al.~\cite{zhang2024extractdefinecanonicalizellmbased} propose using LLMs to construct knowledge graphs from natural language. Fang et al.~\cite{fang2025proofstrategyextractionllms} design a workflow to extract lemmas from the LLM reasoning process. However, these methods cannot be applied to our scenario, as they are designed for obtaining specific types of rules (e.g., logic rules, mathematical equations, or lemmas) from specific data sources (e.g., knowledge graphs, numeric data, or LLM reasoning trajectories), but not for learning tactics from formal proofs. Compared with them, \approach{} uses data-driven guidance to mine logical reasoning strategies formed internally by LLMs, and uses mined tactics enhancing the capabilities of symbolic provers. 
Recent work uses LLMs to extract structured knowledge such as logic rules~\cite{DBLP:conf/pakdd/LuoJXLHP25}, mathematical equations~\cite{DBLP:conf/iclr/ShojaeeMGFR25}, knowledge graphs~\cite{zhang2024extractdefinecanonicalizellmbased}, and lemmas~\cite{fang2025proofstrategyextractionllms}. These methods target specific rule types and data sources, and none of them mine reusable tactics from formal proofs. \approach{} fills this gap by mining reasoning strategies from existing proofs and using the mined tactics to enhance symbolic provers.

The most related work is \stratrocq{}~\cite{fang2025proofstrategyextractionllms}, which also enhances automatic theorem proving but extracts lemmas rather than tactics. \approach{} differs in three ways: (1) we mine reusable tactics from existing formal proofs (written by humans or LLMs), rather than extracting strategies solely from LLM internals; (2) tactics are more generally applicable than lemmas, which only match proof goals of specific types; and (3) we introduce an indirect retrieval method that addresses the lack of textual information in Ltac tactics. Our evaluation shows that \approach{} significantly outperforms \stratrocq{}.
% In particular, the recently published, non-peer-reviewed work \stratrocq{}~\cite{fang2025proofstrategyextractionllms} is most relevant to our work, as \stratrocq{} also enhances automatic theorem proving by extracting lemmas. However, there are several significant differences. First, the goal of \stratrocq{} is to extract the internal proving strategies from LLMs, while our goal is to mine reusable tactics from existing formal proofs. Since the existing formal proofs can be written by either LLMs or human experts, our work can potentially extract strategies that are not present in LLMs. 
% Second, \stratrocq{} focuses on extracting lemmas, while our work focuses on mining tactics. Compared with lemmas which can only match proof goals of specific types, tactics are more generally applicable. 
% For example, perform induction on a variable of any inductive type, or apply a lemma in the context to another term in the context. 
% Third, \approach{} introduces an indirect retrieval method for Ltac tactics that overcomes the lack of textual information in Ltac tactics, whereas \stratrocq{} does not involve any innovation in theorem retrieval.
% As shown in our evaluation, \approach{} significantly outperforms \stratrocq{} in terms of enhancing symbolic provers.

\paragraph{Machine learning-based methods}
Many machine learning methods have been applied to theorem proving~\cite{DBLP:conf/icml/YangD19,DBLP:journals/pacmpl/FirstBG20,DBLP:conf/pldi/Sanchez-SternAS20,DBLP:conf/icml/BlaauwbroekORMP24,DBLP:conf/lpar/BlaauwbroekUG20,DBLP:conf/iclr/WangXZLCHXSX0LL24,DBLP:journals/corr/abs-2410-19940,DBLP:conf/kbse/LuD024,jiang2022thor,DBLP:conf/icse/ThompsonSCFSB0L25}. These approaches can be divided into two categories: i) training new models for theorem proving through large amounts of corpus, and ii) implementing agents to interact with LLMs to generate proof scripts. In other way, a non-trivial neural model is involved in the proving process. On the other hand, \approach{} focuses on enhancing symbolic provers, and as our evaluation shows, the enhancement complements neural models, and can further improve the performance of proving agents as well as reducing their costs.

\paragraph{Symbolic provers}
Many existing symbolic provers are proposed to reduce the burden of manually writing proof scripts, such as CoqHammer~\cite{DBLP:journals/jar/CzajkaK18}, Lean-auto~\cite{qian2025leanautointerfacelean4}, and Sledgehammer~\cite{DBLP:conf/cade/BohmeN10}. These provers typically focus on integrating existing automated theorem provers (ATPs) such as SMT solver to prove theorems in the target proof language, and none of them attempt to mine new tactics from existing proofs as far as we are aware.

% \approach{} only needs to formalize the extracted tactics into the programming language of other provers to enhance other proof systems.

\section{Conclusion and Future Work}
In this paper, we studied how to enhance provers to mine generalizable tactics from large language models. We proposed \approach{}, the first framework for mining tactics from LLMs with support for generalization validation. Our evaluation experiments demonstrate that by using \approach{}, the success rate of CoqHammer improved by \finalres{}.

Currently, because automated theorem provers in interactive theorem assistants do not support user-defined tactics, the mined tactics can only be used externally to the symbolic prover. But this way cannot be deeply integrated into the automated proving processes of the symbolic solvers. How to integrate mined tactics into the automated proving processes remains as future work.

% \section{Data Availability}
% The implementation of \approach{} and all data of this paper are available at \url{https://doi.org/10.5281/zenodo.19251765}.
\bibliographystyle{ACM-Reference-Format}
\bibliography{sample-base}

\end{document}